\documentclass[copyright]{eptcs}
\providecommand{\event}{PLACES 2009} 
\providecommand{\volume}{??}
\providecommand{\anno}{2009}
\providecommand{\eid}{??}
\usepackage{breakurl}        

 \usepackage{space}
 \usepackage{subfigure}
 \usepackage{amssymb}
 \usepackage{amsmath}
\usepackage{listings}
\usepackage[dvipsnames]{xcolor}

\newcommand{\SJ}[1]{\lstinline@#1@}

\title{Session-Based Programming for Parallel Algorithms\\
\Large{Expressiveness and Performance}}
\author{
\quad Andi Bejleri \qquad\qquad Raymond Hu \qquad\qquad Nobuko Yoshida
\institute{Imperial College
London, UK}
\email{\quad ab406@doc.ic.ac.uk \quad\qquad rhu@doc.ic.ac.uk \quad\qquad yoshida@doc.ic.ac.uk}
}
\def\titlerunning{Session-Based Programming for Parallel Algorithms}
\def\authorrunning{A. Bejleri \& R. Hu \& N. Yoshida}
\begin{document}
\maketitle


\begin{abstract}
This paper investigates session programming and typing of benchmark examples to compare productivity, safety and performance with other communications programming languages.
Parallel algorithms are used to examine the above aspects due to their extensive use of message passing for interaction, and their increasing prominence in algorithmic research with the rising availability of hardware resources such as multicore machines and clusters.
We contribute new benchmark results for SJ, an extension of Java for type-safe, binary session programming, against MPJ Express, a Java messaging system based on the
MPI standard. In conclusion, we observe that \emph{(1)} despite rich libraries and functionality, MPI remains a low-level API, and can suffer from commonly perceived disadvantages of explicit message passing such as deadlocks and unexpected message 
types,
and \emph{(2)} the benefits of high-level session abstraction, which has significant impact on program structure to improve readability and reliability, and session type-safety can greatly facilitate the task of communications programming whilst retaining competitive performance.
\end{abstract}

\section{Introduction}
\label{sec:introduction}

At PLACES'08, we discussed the need to investigate benchmark examples
of session types \cite{DBLP:conf/parle/TakeuchiHK94,DBLP:conf/esop/HondaVK98} to compare
productivity, safety and performance with
other communications programming languages.
As a starting point into the investigation of these issues,
we examine SJ \cite{SJwebsite}, the
first full object-oriented language to incorporate session types for
type-safe concurrent and distributed programming. The SJ language
extends Java with syntax for declaring session types (protocols), and
a set of core operations (session initiation, send/receive) and
high-level constructs (branching, iteration, recursion) for
implementing the interactions that comprise the sessions. The SJ
compiler statically verifies session implementations against their
declared types. Together with runtime compatibility validation between
peers at session initiation, SJ guarantees communication safety in
terms of message types and the structure of interaction. SJ has been
shown to perform competitively with widely-used communication APIs
such as network sockets, in certain cases out-performing RMI
\cite{SJecoop08}.


This paper reports our on-going work on implementing parallel
algorithms in SJ,
with focus on the aforementioned aspects:
{\bf\em productivity} (including code readability and writability),
{\bf\em safety}
(freedom from type and communication errors \cite{DBLP:conf/parle/TakeuchiHK94,DBLP:conf/esop/HondaVK98}),
and {\bf\em performance} (optimisations enabled by SJ, and comparison against other communication systems).
Parallel algorithms is a prominent topic in algorithmic research due to the increase of hardware resources such as multicore machines and
clusters.
The session-based programming methodology and
expressiveness of SJ are demonstrated through implementations of: (1)
a Monte Carlo approximation of $\pi$, (2) the Jacobi solution of the
Discrete Poisson Equation, and (3) a simulation of the $n$-Body
problem. These algorithms were selected
to evaluate the SJ representation of, amongst other features, typical {\em task and data decomposition} patterns
\cite{patternsforparallelalgo} (as featured in 1 and 2), a
technique for exchanging {\em ghost points} \cite{usingmpi}
(in 2), and an intricate communication pattern over a circular
pipeline structure (3). SJ is an evolving framework, and recent
extensions
to the
SJ language \cite{SJecoop08}
(e.g. new multicast output operations and advanced iteration structures)
and the SJ Runtime
(e.g.~improved extensibility through the Abstract Transport)
play an important part in the implementation of these algorithms.

Using these programs,
which feature complex
and representative interaction structures, we contribute new benchmark results for analysis
to supplement the existing benchmarks for SJ.
In particular, benchmark comparisons between SJ and MPJ Express \cite{mpjexpress}, a reference Java messaging system based on the MPI \cite{MPI} standard, for (1) and (2) yield further promising performance results for SJ.
We also show how
SJ {\em noalias} types can greatly optimise performance, such as for
the shared memory communication of the ghost points in (2).


We then compare the SJ implementations of the above algorithms
with 
their MPI counterparts from programming perspectives.
Despite rich libraries and functionality, MPI remains a
low-level API, and can suffer from such commonly perceived
disadvantages of explicit message passing as unexpected message
structures and deadlocks due to incorrect protocol implementations.
From our experiences implementing the above algorithms, we found
high-level session programming to be easier than the basic MPI
functions, which often require manipulating numerical process
identifiers and array indexes (e.g.~for message lengths in (3)) in tricky
ways. SJ is able to exploit session types to compensate for, or
eliminate, many of the MPI problems: session types themselves are
inherently deadlock free, for example.

In conclusion, we observe that high-level session abstraction has
significant impact on program structure, improving
readability and reliability, and session type-safety
can greatly facilitate the task of communications programming whilst retaining competitive performance. We also argue that
extending SJ with full multiparty session types would allow richer
topologies such as the ring and 2D-mesh to be expressed more
naturally, and enable performance improvements through massive
parallelism.

\section{Monte Carlo $\pi$ Approximation}
\label{sec:pi}

A simple Monte Carlo simulation for approximating the value of $\pi$
is amenable to parallelisation. We use this example to (1)
introduce basic and some new SJ constructs; (2) show their use in the description of a simple task decomposition pattern
\cite{patternsforparallelalgo}; and (3) demonstrate the effect of parallelisation for performance gain in SJ (\S~\ref{sec:bmarks}).

A unit square inscribes a circle of
area $\pi/4$; hence, $\pi = 4t$, where $t$ is the ratio of the circle
area to the square. $t$ can be determined by selecting a random set of
points within the square ($(x, y)$ where $x, y \in [-1, 1]$), and
checking how many fall inside the inscribed circle ($x^2 + y^2 <=
1$). A Master process (or thread) can instruct Workers to independently
generate and check multiple sets of points in parallel, calculating
the final value by combining the results from each Worker. The simple
session type, from the Worker side, for the communications
involved is:

\begin{center}
\SJ{protocol workerToMaster \{ sbegin.?(int).!<int> \}}
\end{center}
Each Worker service (\SJ{sbegin}) is told how many points to test by the Master (\SJ{?(int)}) and sends back the number that fall inside the circle (\SJ{!<int>}). The code for a basic SJ implementation looks like

\vspace{1mm}
\noindent
\begin{tabular}{ll}
\begin{lstlisting}
// Workers run the simulation.
int trials = s_wm.receive(); // ?(int)
for(int i = 0; i < trials; i++)
    if(hit()) hits++;					
s_wm.send(hits); // !<int>
\end{lstlisting}
&
\begin{lstlisting}
// Master controls the Workers.
<s_mw1, s_mw2, ...>.send(trials); // Multicast.
int totalHits = // Collect the results.
    s_mw1.receive()
      + s_mw2.receive() + ...;
\end{lstlisting}
\end{tabular} \\
\vspace{1mm}

\noindent
where \SJ{s_mw1} is the Master's session socket to Worker1, etc.; \SJ{s_wm} a Worker's session with the Master; and \SJ{hit} returns the boolean from testing a generated point. The Master can then calculate $t$ by \SJ{totalHits / (trials * n)}, where \SJ{n} is the number of Workers.

The SJ compiler statically verifies correctness by checking each session implementation against its declared type (e.g. \SJ{s_wm} against \SJ{workerToMaster}). Then at runtime, session initiation validates the session types of each peer to ensure \emph{duality} between the peers. If successful, the
session is established; otherwise, both parties raise an
\SJ{SJIncompatibleSessionEception} and the session
is aborted. The SJ Runtime is also responsible for failure handling during session execution: if an error occurs at one session peer, e.g. an exception is raised, the failure signal is propagated to all relevant session parties, maintaining consistency across dependent sessions;
see \cite{SJecoop08} for more detailed explanation.

\section{Jacobi Solution of the Discrete Poisson Equation}
\label{sec:jacobi}

The implementation of this algorithm demonstrates (1) the expressiveness
of SJ due to multicast session-iteration operation;
(2) guaranteed type and communication safety in SJ; (3) a
type-directed optimisation (for exchanging ghost points) using the new
SJ {\em noalias} type; and
(4) the {\em transport-independence} of SJ programs, due to the design of the SJ language-runtime framework. Poisson's Equation is a partial differential equation with
applications in, for example, heat flow, electrostatics, gravity and
climate computations.
The
discrete two-dimensional Poisson equation $(\nabla^2u)_{ij}$ for a $m \times n$ grid
can be written as the formula in (a),
%
%

\vspace{4mm}
\begin{figure}[h]
\centering

(a)~ $u_{ij} =
		\frac{1}{4}(u_{i-1, j} + u_{i+1, j} + u_{i, j-1} + u_{i, j+1} - dx^2g_{i, j})$
%
\hspace{6mm}
%
(b)~ $u_{ij}^{k+1} =
	\frac{1}{4}(u_{i+1, j}^k + u_{i-1, j}^k + u_{i,j+1}^k + u_{i, j-1}^k)$


\label{fig:jacobi}

\vspace{4mm}
\end{figure}

\noindent
where $2 \leq i \leq m - 1$, $2 \leq j \leq n - 1$, and $dx = 1 / (n + 1)$. Jacobi's Method converges on a solution by repeatedly replacing each element of the matrix $u$ by an
average of its four neighbouring values and $dx^2g_{i, j}$. For this example, we set $g$ to $0$; then from the $k$-th approximation of $u$, the next iteration
performs the calculation in (b) above.
%
%
%
%
%
%
\noindent Termination may be on reaching a target convergence threshold or completing a certain number of iterations. Parallelization exploits the fact that each element can be updated independently (within one step): the grid can be divided up and the algorithm performed on each subgrid in separate processes or threads. The key is that neighbouring processes must exchange their subgrid boundary values as they are updated.

We illustrate a one-dimensional decomposition of a square grid into three non-overlapping subgrids for three separate processes. Two Workers are allocated the end subgrids; the Master has the central subgrid, and controls the termination condition for all three processes. In addition to their allocated subgrid, each process maintains a copy of the boundary values ({\em ghost points}) of its neighbours; the new values are communicated after each iteration. This scheme allows the original grid to be divided in subgrids of any size. The session type between the Master and each of the two Workers from the side of the former is:

\vspace{1mm}
\noindent
\begin{lstlisting}
protocol masterToWorker {
  cbegin.                        // Request the Worker service.
  !<int>.                        // Send the size of the matrix.
  ![                             // Enter the main loop (check termination condition).
    !<double[]>.?(double[]).     /* Send our boundary values and..
                                    ..get the Worker's updated ghost points. */
    ?(double).?(double)          // Receive the convergence data for Worker's subgrid.
  ]*.                            // After the last iteration..
  ?(double[][])                  // ..get the final results.
}
\end{lstlisting}
\vspace{1mm}

To control all the Workers simultaneously, the implementation of Master uses the SJ session constructs for multicasting output operations such as message-send and also session-iteration (see Appendix~\ref{app} for the full implementation). For example:

\vspace{1mm}
\noindent \hspace{1mm}
\begin{tabular}{lp{4mm}l}
\begin{lstlisting}
// Master controls iteration condition.
<mw1, mw2>.outwhile(   // ![..
    !accurateEnough(...) && iters < MAX_ITERS) {
  ... // Main body of the algorithm.
}                      // ..]*
\end{lstlisting}
&&
\begin{lstlisting}
// Workers obey the Master.
<wm>.inwhile() {   // ?[..
  ... /* Main body of
          the algorithm. */
}                  // ..]*
\end{lstlisting}
\end{tabular} \\
\vspace{1mm}

\noindent Like the standard while-statement, the outwhile operation evaluates the boolean
condition for iteration (\SJ{!accurateEnough(...)} \SJ{\&\& iters $<$ MAX\_ITERS}) to determine whether the loop continues or terminates. The key difference is that this decision is implicitly communicated to the session peer (in this case, from Master to the two Worker), synchronising the
control flow between two parties. Worker is programmed with the dual behaviour: \SJ{inwhile} does not specify a
loop-condition because this decision is made by Master and communicated to
Worker at each iteration.

Inter-thread communication of large messages, such as arrays, can be optimised using SJ \SJ{noalias} types. A \SJ{noalias} variable on the RHS of an assignment or as a method argument --- such as to the \SJ{send} operation --- becomes \SJ{null} after the assignment or the method call. Combined with static type checking that precludes any potential assignment of aliased values to \SJ{noalias} targets, a \SJ{noalias} variable is guaranteed the sole reference to the pointed object at all times, permitting zero-copy message passing of \SJ{noalias} messages over compatible shared memory transports. In the present example, the \SJ{noalias} optimisation can be used to communicate the ghost point data; for example, the Worker implementations contain the following code extract.

\vspace{1mm}
\noindent
\begin{lstlisting}
// noalias array containing our boundary values (ghost points for the Master).
noalias double[] ghostPoints = ...; /* Update and prepare our boundary values
                                           for sending. */
s_wm.send(ghostPoints); // Type-directed zero-copy send: !<noalias double[]>
... // ghostPoints variable becomes null.
\end{lstlisting}
\vspace{1mm}

\noindent Transports that do not support this feature (e.g. TCP) can fall back to copy-on-send; the overall semantics of the program remains unchanged. This illustrates the {\em transport-independent} nature of SJ programs: the virtualisation of communication due to the SJ Runtime allows programs to make the best use of the whichever transports are available, {\em without} requiring any modification to the programs themselves. If the Master and Worker processes are run on separate machines, then the SJ Runtime can arrange, e.g. a TCP-based session; for the same programs, run as co-located threads, shared memory will be used. This SJ feature is further demonstrated for the next algorithm.

\section{The $n$-Body Problem}
\label{sec:nbody}


The $n$-Body Problem involves finding the motion, according to
classical mechanics, of a system of bodies given their masses and
initial position and velocities.
This advanced example demonstrates (1) the expressiveness of SJ and the extensions for complex iteration structures,
by implementing an intricate
circular communication pipeline; (2) SJ transport-independence (see \S~\ref{sec:bmarks}); and (3) the benefits of high-level message types (see \S~\ref{sec:conc}).
Parallelism is achieved by dividing the particle set, and hence the calculations to determine the resultant force exerted on each body, amongst a collection of parallel processes.
We use the approach where the processes, maintaining only the current state of their individual particle sets, are deployed to form a
circular pipeline (ring topology). Firstly, the number
of processes in the pipeline, $p$, is dynamically determined by
sending a token around the ring.
Then each step of the simulation
involves $p-1$ iterations. In the first iteration, each process sends
their particle data to their neighbour on the right and calculates the
partial resultant forces exerted within their own particle set. In the
n-th iteration, each process forwards on the particle data received in
the previous iteration (line~({\em i}) in Figure~\ref{fig:mpi-sj2}), adds this data to the running force
calculation ({\em ii}), and receives the next data set ({\em iii}). The particle data from the right neighbour is received by
the end of the final iteration: each data set has now been seen by all
processors in the pipeline, allowing the final results for the current
simulation step to be calculated.

The SJ implementation of the above algorithm has each process, i.e. each Worker unit in the pipeline, open a
session server socket to accept a connection from its left neighbour,
and create the connection to its right neighbour using a session
client socket. The session type for the
interaction in this algorithm, from the server side of each unit, is:

\vspace{1mm}
\noindent \hspace{1mm}
\begin{tabular}{ll}
\begin{lstlisting}
protocol serverSide {
  sbegin.
  !<int>.
  ?[
    ?[
      ?(Particle[])
    ]*
  ]*
}
\end{lstlisting}
&
\begin{lstlisting}
// Interaction with the left neighbour.
// Accept connection from left neighbour.
// Forward on the ring initialisation token.
// Main simulation loop (iteration flag received from the left).
// Inner iterations within each simulation step.
// Particle data forwarded through pipeline.
(*@@*)
(*@@*)
(*@@*)
\end{lstlisting}
\end{tabular} \\
\vspace{1mm}

\noindent
The session type for the corresponding client side of each unit is simply the direct dual of \SJ{serverSide}: \SJ{protocol clientSide \{ cbegin.?(int).![![!<Particle[]>]*]*} \}, given by inverting the input (\SJ{?}) and output (\SJ{!}) symbols. For this client-server architecture, the ring topology
is bootstrapped by designating two neighbouring processes to be the
``first'' and ``last'' pipeline units.

The remaining SJ code for this example and a comparison with an MPI implementation (Figure~\ref{fig:mpi-sj2}) are outlined in \S~\ref{sec:conc}.

\section{Performance Benchmarks}
\label{sec:bmarks}

This section presents performance measurements
for the three parallel algorithms described above.
The first two benchmarks show that the SJ Runtime, although still at an early implementation version with much scope for further optimisation, can perform competitively with MPJ Express \cite{mpjexpress}.
Unlike Java MPI implementations built around JNI wrappers to C functions, MPJ Express adopts a pure Java approach which makes for
a more informative comparison with SJ.

The same
machines in the same network environment were used for all the following
benchmark experiments. Each machine is a dual-core Intel Core 2
Duo (Conroe~B2) at 2.13GHz with 2MB cache, 2GB main memory,
running Ubuntu Linux~4.2.3 (kernel~2.6.24); the machines were
connected via gigabit Ethernet, and the latency between two machines
was measured using ping (64~Bytes) to be on average 0.10ms. The benchmark applications were
compiled and executed using the standard Sun Java SE compiler and
runtime versions~1.6.0. For each experiment, the results from 100 executions for each parameter configuration were recorded; here, we give the mean values. The full source code for the benchmark applications and the complete results can be found at \cite{parallelalgowebsite}.

\paragraph{Monte Carlo $\pi$ approximation.} The first benchmark uses the SJ implementation of this algorithm to (1) verify the performance gain from increased parallelism, and (2) to compare the performance of the SJ Runtime against MPJ Express. Each process (Master, Workers and Client) was run on a separate machine,
communicating via TCP. The results (Figure~\ref{fig:benchmarks12}), comparing both sequential and parallel versions of the algorithm, show that for a constant sample size (total number of test points), increasing the number of Workers indeed reduces the time to complete the algorithm proportionally. The results for the SJ implementation are around 5--6\% faster than the MPJ Express implementation.

\begin{figure}[t]
\centering

	\begin{tabular}{|l||rr|}
		\hline
		\textbf{Configuration} & SJ (ms) & MPJ (ms) \\
		\hline
		Sequential (1 Worker) & \multicolumn{2}{c|}{6717} \\
	 \hline
		1 Master \& 1 Worker & 3764 & 3846 \\
	 \hline
	 	1 Master \& 2 Workers & 2466 & 2606 \\
	\hline
		1 Master \& 3 Workers & 1885 & 1966 \\
	\hline
		1 Master \& 4 Workers & 1487 &  1579\\
		\hline
	\end{tabular}

\caption{ Monte Carlo $\pi$ for a varying number of Workers.}
\label{fig:benchmarks12}

\rule{\linewidth}{0.2mm}
\end{figure}

\paragraph{Jacobi Poisson solution.} The second benchmark, through the SJ implementation of the Jacobi iteration algorithm,
demonstrates (1) the effectiveness of \SJ{noalias} types for zero-copy
message transfer in a shared memory environment, and (2) again compares SJ performance to MPJ Express.
Firstly, ``Ordinary'' (i.e. without \SJ{noalias}) and \SJ{noalias} versions of the Master
and two Workers were run as co-VM threads on a single machine; the Client is connected to
the Master from a separate machine via a TCP-session. We measured the time to complete the
algorithm for square matrices of size (i.e. the length of one side of the matrix) 100 and 300. In both cases, the \SJ{noalias} version is approximately 20\%
faster than the ordinary one (Figure~\ref{fig:benchmarks3}). For sizes greater than 300, we observed that the local computation costs start to dominate the
communication costs for this fixed number of Workers, reducing the differences between the execution times of the ``Ordinary" and \SJ{noalias} versions, e.g. for matrix size 1000.
Secondly, the distributed SJ
implementation of Jacobi (the Client, Master and Workers run on separate
machines connected via TCP) performs better than the MPJ Express implementation by 6\% on  average (Figure~\ref{fig:benchmarks4}).\\

\begin{figure}[t]
\centering
\vspace{1mm}

\subfigure[]
{
	\begin{tabular}{|l||rr|}
		\hline
		\textbf{Matrix Size} & ``Ordinary'' (ms) & \SJ{noalias} (ms) \\
		\hline
		100 & 1270 & 992 \\
		\hline
		300 & 24436 & 19448 \\
		\hline
		1000 & 288532  & 299279 \\
		\hline
	\end{tabular}
	
\label{fig:benchmarks3}
}
\hspace{5mm}
\subfigure[]{

	\begin{tabular}{|l||rr|}
		\hline
	 \textbf{Matrix Size} & SJ (ms) & MPJ (ms) \\
		\hline
		100 & 3713 & 4460 \\
	 \hline
		300 & 19501 & 19834\\
	 \hline
	\end{tabular}
	
\label{fig:benchmarks4}
}

\vspace{-2mm}

\caption{(a) Jacobi: ``ordinary'' vs. \SJ{noalias} versions; (b) Jacobi: SJ vs. MPJ Express.}
\label{fig:benchmark3}
\rule{\linewidth}{0.2mm}
\end{figure}

\paragraph{$n$-Body simulation.} The third benchmark uses the $n$-Body simulation to demonstrate the important improvement in
productivity enabled by SJ transport-independence: this single SJ implementation was run in the different
communication environments (locally concurrent, distributed), making the best use of the available transports (TCP, shared memory, etc.), without {\em any} changes to the source
code for the Workers (although the shared memory version required a few lines of external code to bootstrap the Workers as Java threads). The benchmark was executed using two pipeline Worker units (not using \SJ{noalias}) in three different configurations: the two Workers on separate machines using TCP (Distributed),
as separate processes on the same machine using TCP (Localhost), and as co-VM threads using shared memory (Threads). We recorded the results for simulations involving 100, 300 and 1000 particles, distributed equally between the Workers.

As expected, the results (Figure~\ref{fig:benchmark4}) show the Threads version is faster than
Localhost: around 27\% for 100 particles, 24\% for 300, and 10\% for 1000. The Distributed version is
in turn slightly slower (latency is very low) than Localhost: 10\% for 100 particles, 4\% for 300, and 3\% for 1000.
The relative performance gain between each version decreases for larger particle sets because the local computation costs begin to dominate the communication costs for this fixed number of Workers. Naturally, performance can be improved for simulations involving many particles by increasing the degree of parallelism, i.e. using more Workers.\\


\begin{figure}[t]
\centering
\begin{tabular}{|l| |rrr|}
		\hline
	 \textbf{Particles} & Distrib. (ms) & Localhost (ms) & Threads (ms) \\
		\hline
		100 & 496 & 452 & 326\\
	 \hline
		300 & 1194 & 1144 & 865 \\
	 \hline
		1000 & 7702 & 7497 & 6785\\
		\hline
	\end{tabular}
	
\caption{$n$-Body simulation: Distributed vs. Localhost vs. Threads versions.}
\label{fig:benchmark4}

\rule{\linewidth}{0.2mm}
\end{figure}


\section{SJ and MPI Comparison} 
\label{sec:conc}




This section compares SJ against MPI in terms of language support for
communications programming, with reference to MPI implementations
of the above algorithms \cite{usingmpi}.
Since MPI has an extensive library of functions developed over 15 years, many of these are not yet directly supported
in SJ, e.g. 
MPI Jacobi makes use of a virtual topology
(\SJ{MPI\_Cart\_Create}) and collective data movement operations
(\SJ{MPI\_Bcast} and \SJ{MPI\_Allreduce}, for
broadcasting the matrix size and distributing the termination condition in (2)). However, many of these features can be encoded into a session type, as shown above. Furthermore, we observed the following benefits of SJ against MPI.

\paragraph{Type and communication safety from session types.}
MPI is designed as a portable API specification to be implemented for
varying host languages. Coupled to the low-level nature of
many MPI functions, the design of accompanying MPI program
verification techniques for a host language can be difficult. Common
MPI errors recognized by the community include:

\begin{itemize}
\setlength{\topsep}{0mm}
\setlength{\leftmargin}{0mm}
\setlength{\itemindent}{0mm}
\setlength{\itemsep}{0mm}
\setlength{\listparindent}{0mm}

\item{{\bf Invalid actions before \SJ{MPI\_Init} and after \SJ{MPI\_Finalize}}.}
	The execution of such MPI operations can lead to runtime errors such as broken invariants, messages not broadcasted,
	and incorrect collective operations. Figure~\ref{fig:mpi-sj} presents the correct code of setting up the topology in the $n$-body simulation in MPI\footnote{This MPI implementation of the $n$-Body simulation
is taken from the Using MPI website \cite{MPI}.} (left column) and SJ (right column). In the MPI code, the errors we are referring to would come from adding MPI operations before line~\ref{line:mpi1} and after line~\ref{line:mpi2}. In SJ, actions incorrectly performed before the server socket (line~\ref{line:sj1}) or the session (lines~\ref{line:sj2}--\ref{line:sj3}) have been initialised are rejected by the compiler. The static type system of SJ also does not allow session actions to be performed after leaving the relevant session-try scope (i.e. on \SJ{left} or \SJ{right} after line~\ref{line:sj4}). The MPI and SJ code for the main body of the algorithm is given in Figure~\ref{fig:mpi-sj2}.

\item{{\bf Unmatched \SJ{MPI\_Send} and \SJ{MPI\_Recv}}.}
	Such errors can lead to a mismatch between the sent and expected
message type/structure, or a variety of deadlock situations depending
on the communication mode. For example, two processes deadlock if each
is waiting for a message before sending the message expected by the
other. In the standard (buffer-blocking) mode, the converse situation
(both processes attempting to send before receiving) can also
deadlock: if both message sizes are bigger than the available space in
the medium and opposing receive buffers, then the processes cannot
complete their write operations. A related problem is matching a \SJ{MPI\_Bcast} output with \SJ{MPI\_Recv}. Standard usage is to receive a broadcast message using the complementary \SJ{MPI\_Bcast} input. \SJ{MPI\_Recv} consumes the message; hence, the receiver must be able to determine which processes have not yet seen the message and manually re-broadcast it.

\item{{\bf Concurrency issues}.}
	Incorrect access of a shared communicator by separate threads
can violate the intended message causalities between the
sender(s) and the receivers. In addition, race conditions can
arise due to modifying, or even just by accessing, messages that are in
transit.
\end{itemize}

\begin{figure}[t]
{\lstset{
basicstyle=\small, 
numbers=left, numberstyle=\tiny, stepnumber=1, numbersep=5pt
}
\noindent
\begin{tabular}{ll}
\begin{lstlisting}
main(int argc, char *argv[]) {
  // Set up of the topology.
  MPI_Init(&argc, &argv); (*@\label{line:mpi1}@*)
  MPI_Comm_rank(MPI_COMM_WORLD, &rank);
  MPI_Comm_size(MPI_COMM_WORLD, &size);
  // Get the best ring in the topology.
  periodic = 1;
  MPI_Cart_create(MPI_COMM_WORLD, 1,
        &size, &periodic, 1, &commring);
  MPI_Cart_shift(commring, 0, 1,
        &left, &right);
  ... // Main algorithm body.
  MPI_Finalize(); (*@\label{line:mpi2}@*)
  return 0;
}
(*@~@*)
(*@~@*)
(*@~@*)
 \end{lstlisting}
&
\begin{lstlisting}
public void run(...) {
  // Set up the sockets for the topology.
  SJService c_r =
      SJService.create(pc_nbody, host_r, port_r);
  SJServerSocket ss_l;
  SJSocket left, right;
  try(ss_l) {
    ss_l = SJServerSocket.create(ps_nbody, port_l); (*@\label{line:sj1}@*)
    try(left, right) {
      left = ss_l.accept(); (*@\label{line:sj2}@*)
      right = c_r.request(); (*@\label{line:sj3}@*)
      // Determine the topology size.
      left.send(right.receiveInt() + 1);
      ... // Main algorithm body.
    } finally {...} (*@\label{line:sj4}@*)
  } catch(SJIncompatibleSessionException ise) {...}
  ...// Handling for other exceptions.
}
\end{lstlisting}
\end{tabular}}

\caption{Setting up the topology for the $n$-Body simulation in MPI and in SJ.}
\label{fig:mpi-sj}

\rule{\linewidth}{0.2mm}
\end{figure}

\noindent
As illustrated in the previous sections, {\bf\em SJ programs are guaranteed free from all of the above errors} by the semantics of session communication and static session type
checking.
The first two points are directly prevented by the
properties of session types. For the third point, the SJ compiler
disallows sharing of session socket objects (implicitly \SJ{noalias}), and message copying/linear transfer can
be safely and explicitly controlled via \SJ{noalias} types.

\paragraph*{High-level message types.}
In many parallel algorithms, messages are mainly communicated via
arrays. For MPI, effort is required to manually track and communicate
array indices, e.g.~for message length or the number of messages. In
contrast, the high-level type-abstraction for messages allows SJ programmers to treat both object and primitive array
messages as regular Java array objects. For instance, the MPI version of the main algorithm for the $n$-Body simulation\footnotemark[1] (Figure~\ref{fig:mpi-sj2}, left) broadcasts the number of particles
managed by each
process, through the \SJ{MPI_Allgather} operation (line~\ref{line:mpi21}). Thus, the amount of data to be read from each particle set (line~\ref{line:mpi22})
can be determined (lines~\ref{line:mpi23}--\ref{line:mpi24}). In SJ (Figure~\ref{fig:mpi-sj2}, right), the particle data is simply received as discrete array messages (line~\ref{line:sj21}), avoiding manual handling of
message sizes. Therefore, the MPI code between lines~\ref{line:mpi21}--\ref{line:mpi24} is unnecessary in the SJ implementation. The rest of the code structure is the same in both implementations.

In the SJ implementation of the $n$-Body, the assignment in (\emph{iii}) is permitted because the received message is implicitly \SJ{noalias}.

\begin{figure}[t!]
{\lstset{
basicstyle=\small, 
numbers=left, numberstyle=\tiny, stepnumber=1, numbersep=5pt
}
\noindent \begin{tabular}{ll}
\begin{lstlisting}
// Get the sizes and displacements.
MPI_Allgather(&npart, 1, MPI_INT, counts, (*@\label{line:mpi21}@*)
      1, MPI_INT, commring);
displs[0] = 0; (*@\label{line:mpi23}@*)
for(i=1; i<size; i++)
  displs[i] = displs[i-1] + counts[i-1];
totpart = displs[size-1] + counts[size-1]; (*@\label{line:mpi24}@*)
InitParticles(particles, pv, npart);
while(cnt--) {
  double max_f, max_f_seg;
  // Load the initial sendbuffer.
  memcpy(sendbuf, particles,
        npart * sizeof(Particle));
  for(pipe=0; pipe<size; pipe++) {
    if(pipe != size-1) {
      MPI_Isend(sendbuf, npart, particletype,
           right, pipe, commring, &request[0]);
      MPI_Irecv(recvbuf, npart, particletype, (*@\label{line:mpi22}@*)
           left,  pipe, commring, &request[1]);
    }
    // Compute forces.
    max_f_seg = ComputeForces(particles,
                      sendbuf, pv, npart);
    // Wait for non-blocking receives to return.
    if(pipe != size-1)
      MPI_Waitall(2, request, statuses);
    memcpy(sendbuf, recvbuf,
          counts[pipe] * sizeof(Particle));
  }
  // Update our own particle data.
  sim_t += ComputeNewPos(particles, pv, npart,
                max_f, commring);
}
\end{lstlisting}
&\quad
\begin{lstlisting}
initParticles(particles, pvs);
/* Synchronise with our two neighbours
   for each simulation step. */
right.outwhile(left.inwhile()) {
  // Load the initial sendbuffer.
  Particle[] current =
        new Particle[numParticles];
  System.arraycopy(particles, 0, current,
        0, numParticles);
  /* Inner iterations within each
     simulation step. */
  right.outwhile(left.inwhile()) {
    // (*@(\emph{i})@*) Forward the current data set.
    right.send(current);
    /* (*@(\emph{ii})@*) Add the current data to
       the running calculation. */
    computeForces(particles, current, pvs);
    // (*@(\emph{iii})@*) Receive the next data set.
    current = (Particle[]) left.receive(); (*@\label{line:sj21}@*)
  }
  /* Calculate the final results for
     this simulation step and update
     our own particle data. */
  computeForces(particles, current, pvs);
  computeNewPos(particles, pvs, i);
  i++;
}
\end{lstlisting}
\end{tabular}}

\caption{Implementing the main body of the $n$-Body simulation algorithm in MPI and SJ.}
\label{fig:mpi-sj2}

\rule{\linewidth}{0.2mm}
\end{figure}

\paragraph{Transparent zero-copy message passing.}
SJ provides direct language support for zero-copy transfer in shared
memory contexts through \SJ{noalias} types. This feature can enable
significant performance increases for multi-threaded programs (see
\S~\ref{sec:bmarks}). Moreover, the communication of \SJ{noalias}
types retains consistent semantics in all transport contexts
(see {\em transport-independence} in \S~\ref{sec:jacobi}).


\section{Conclusions and Future Work}
We demonstrated expressiveness, productivity and
performance benefits of session-based programming in SJ through the presented parallel algorithm implementations. Although we have seen that the above algorithms were readily implemented in the current SJ, immediate future work
includes expanding the set of SJ operations and constructs, e.g. with session typed equivalents of MPI functions and features that are not yet directly supported. For example,
whilst the MPI {\em standard} mode (send and receive block on their
respective buffers) corresponds to the session communication semantics
in SJ, MPI has several additional modes: {\em synchronous} (send and
receive operations synchronise), {\em ready} (programmer notifies the
system that a receive has been posted), and {\em buffered} (user
manually handles send buffers).
We also wish to
compare SJ to
PGAS languages such as X10 \cite{X10Homepage} using parallel algorithm implementation as a basis.

We believe that extending SJ with full multiparty session types
\cite{DBLP:conf/popl/HondaYC08} would allow richer
topologies such as the ring and 2D-mesh to be expressed more
naturally in a type-safe manner. For example,
the SJ $n$-Body implementation currently
requires creating one intermediary session (for the final pipeline link) in each simulation step; with multiparty sessions,
we would only need to open a single session for the complete simulation.
Our prediction is that multiparty sessions will offer better support for massive
parallelism than the current client-server based session sockets.
We plan to identify design issues
and possible overheads
for global type-checking through
further implementation of parallel algorithms with complex
communication patterns.

SJ programs are guaranteed free from type and communication errors,
and perform competitively against other Java communication runtimes. In certain
cases, SJ programs can out-perform their counterparts implemented in communication-safe
systems such as RMI \cite{SJecoop08} and also lower-level, non communication-safe message passing systems
such as MPJ Express (\S~\ref{sec:bmarks}).

\section{Acknowledgments}
\label{sec:ack}

We thank the reviewers for their helpful comments
on the submission. We also thank
Kohei Honda and Vijay Saraswat for their comments on a first
draft of this paper.
The work is partially supported by EPSRC GR/T03208 and GR/T03215.

\bibliographystyle{eptcs}
\bibliography{bibliography}

\appendix

\section{Appendix}
\label{app}

The full SJ source code for the Master party of the Jacobi iteration example (\S~\ref{sec:jacobi}) is listed below. SJ \SJ{protocol}s are implicitly \SJ{final} and \SJ{noalias}. We include explicit casts of received messages for clarity; however, this type information can be inferred by the SJ compiler from the declared protocols.
The implementation of the Worker parties can be found at \cite{parallelalgowebsite}.

\begin{lstlisting}
package onedimjacobi.noaliaz;

import java.io.*;
import java.util.*;
import sessionj.runtime.*;
import sessionj.runtime.net.*;

public class Master {
  protocol p_mc sbegin.?(int).!<double[][]> // Master-to-Client.

  protocol matrix_size !<int>
  protocol stopping_condition ?(Double).?(Double)
  protocol ghost_points !<double[]>.?(double[])
  protocol partial_result ?(double[][])

  protocol p_mw { // Master-to-Workers.
    cbegin
    .@(matrix_size)
    .![
      @(ghost_points)
      .@(stopping_condition)
    ]*
    .@(partial_result)
  }

  private static final int MAX_ITERATIONS = 100000;

  public void run(int port_m, String host_n, int port_n,
        String host_s, int port_s)  {
    final noalias SJServerSocket ss; // Server socket for Client requests.

    // Channels for requesting the Worker services (called N and S).
    final noalias SJService c_n = SJService.create(p_mw, host_n, port_n);
    final noalias SJService c_s = SJService.create(p_mw, host_s, port_s);

    try(ss) {
      ss = SJServerSocket.create(p_mc, port_m); // Init. server socket.

      while(true) {
        final noalias SJSocket cm;

        try(cm) {
          cm = ss.accept(); // Accept the Client session request.

          int size = cm.receiveInt(); // The problem size.
          int rows = size / 3;

          final noalias SJSocket mn, ms;

          try(cm, mn, ms) {
            // Set up the Worker sessions.
					  mn = c_n.request();
					  ms = c_s.request();

            <mn, ms>.send(size); // Tell the Workers the problem size.

            // Create the Master's sub-grids for the current and next iterations.
            double[][] u = new double[rows + 2][size + 2];
            double[][] newu = new double[rows + 2][size + 2];

            init(u, newu, rows, size); // Initialise u and newu.

            double diff = 1.0;
            double valmx = 1.0;
            int iterations = 1;

            // Master controls the iteration (termination) condition.
            <mn, ms>.outwhile((diff / valmx) >= (1.0 * Math.pow(10, -5))
                  && iterations <= MAX_ITERATIONS) {
              // Main body of the algorithm.

              diff = 0.0;
              valmx = 0.0;

              // Jacobi iterations.
              for(int i = 1; i < rows + 1; i++) {
                for(int j = 1; j < size + 1; j++) {
                  newu[i][j] = (u[i - 1][j] + u[i + 1][j]
                                     + u[i][j - 1] + u[i][j + 1]) / 4.0;

                  diff = Math.max(diff, Math.abs(newu[i][j] - u[i][j]));
                  valmx = Math.max(valmx, Math.abs(newu[i][j]));
                }
              }

              // Ghost points for the Workers.
              noalias double[] border_n = new double[size];
              noalias double[] border_s = new double[size];

              for(int k = 0; k < size; k++) border_n[k] = newu[1][k + 1];
              for(int k = 0; k < size; k++) border_s[k] = newu[rows][k + 1];

              mn.send(border_n);
              ms.send(border_s);

              // Receive our ghost points from the Workers.
              noalias double[] ghost_n = (double[]) mn.receive();
              noalias double[] ghost_s = (double[]) ms.receive();

              // Copy ghost zones in newu
              for(int k = 0; k < ghost_n.length; k++)
                newu[0][k + 1] = ghost_n[k];
              for(int k = 0; k <  ghost_s.length; k++)
                newu[rows + 1][k+1] = ghost_s[k];

              // Update u with newu.
              double[][] tmp = u;
              u = newu;
              newu = tmp;

              // Computing the new error values.
              diff = Math.max(diff, ((Double) mn.receive()).doubleValue());
              valmx = Math.max(valmx, ((Double) mn.receive()).doubleValue());

              diff = Math.max(diff, ((Double) ms.receive()).doubleValue());
              valmx = Math.max(valmx, ((Double) ms.receive()).doubleValue());

              if(iterations == 1)  {
                diff = 1.0;
                valmx = 1.0;
              }

              iterations++;
            }

            double[][] w1 = (double[][]) mn.receive();
            double[][] w2 = (double[][]) ms.receive();

            double[][] result =  new double[size][size];

            for(int i = 0; i < rows; i++)
              for(int j = 0; j < size; j++)
                result[i][j] = w1[i + 1][j + 1];

            for(int i = rows; i < 2 * rows; i++)
              for(int j = 0; j < size; j++)
                result[i][j] = u[i - rows + 1][j + 1];

            for(int i = 2 * rows; i < size; i++)
              for(int j = 0; j < size; j++)
                result[i][j] = w2[i - 2 * rows + 1][j + 1];

            cm.send(result);
          }
          finally { }
        }
        finally { }
      }
    }
    catch(SJIncompatibleSessionException ise) {
      System.err.println("Incompatible Client type: "+ ise);
    }
    catch(SJIOException sioe) {
      System.err.println("I/O error: " + sioe);
    }
    finally { }
  }
}
\end{lstlisting}


\end{document}